\documentclass[aps,twocolumn,groupedaddress, showpacs,prb]{revtex4}
\usepackage{graphicx}%

\begin{document}
% You should use BibTeX and apsrev.bst for references
\bibliographystyle{apsrev}

%\preprint{}

\title{Making sense of nanocrystal lattice fringes}
% Optional argument for running titles on pages
%\title[]{}


% other informatio
\author{P. Fraundorf}
\affiliation{Dept. of Physics \& Astronomy and Center for Molecular Electronics, U. Missouri-StL, St. Louis MO 63121 USA}
\email[]{pfraundorf@umsl.edu}
\author{Wentao Qin}
\affiliation{Technology Solutions, Freescale Semiconductor, Inc., MD CH305, Chandler AZ 85284 USA}
\author{Peter Moeck}
\affiliation{Dept. of Physics, Portland State University, P. O. Box 751, Portland OR 97207-0751 USA}
\author{Eric Mandell}
%\thanks{}
\affiliation{Dept. of Physics and Astronomy and Center for Molecular Electronics, U. Missouri-StL, St. Louis MO 63121 USA}


\date{\today}

\begin{abstract}
% Text of abstract
The orientation-dependence of thin-crystal lattice fringes
can be gracefully quantified using fringe-visibility maps, a 
direct-space analog of Kikuchi maps. As in navigation 
of reciprocal space with the aid of Kikuchi lines, fringe-visibility 
maps facilitate acquisition of 3D crystallographic information in 
lattice images.  In particular, these maps can help researchers 
to determine the 3D lattice parameters of individual nano-crystals, 
to ``fringe fingerprint'' collections of randomly-oriented particles, 
and to measure local specimen-thickness with only modest tilt.  
Since the number of fringes in an image increases with maximum 
spatial-frequency squared, these strategies (with 
help from more precise goniometers) will be more useful 
as aberration-correction moves resolutions into the subangstrom 
range.  %(Adapted from {\em cond-mat}/{\bf 0212281})

\end{abstract}

%\keywords{}
\maketitle
% main text
% main text

\tableofcontents

\section{Introduction}
\label{sec:Intro}

Nanocrystalline materials have attracted enormous interest due to physical and chemical properties that differ from those of the bulk forms thanks to proper engineering at the atomic level \citep{Kung99,Pitkethly03}.  
Phase diagrams and crystal morphologies are frequently dependent on the size of the crystals in the one to ten nanometer size range 
\citep{Jesser99, Wautelet92, Wautelet01}. In addition to this size-dependence for the lowest energy state of a structure, there is a strong tendency in the nanoparticle regime to metastability \citep{Tolles00} and non-stoichiometry.  Thus needs for characterization of individual nanocrystals will grow.

A simple approach to determining the lattice 
(not atomic) structure of nanocrystals, as though they 
are hand specimens with visible atom columns, was proposed in the 
late 1980's \citep{Phil27}. It relies on the 3D 
reconstruction of lattice periodicities from images 
taken at different tilts.  As in diffraction-based techniques 
\citep{Unwi16, GoehnerEtAl, Bonnet78, KeXin78, Phil7, Phil4, Cowl10} 
for obtaining 3D crystallographic information, and image-based techniques 
in stereomicroscopy \citep{Basi12, Liu2, Liu5, Liu1, Moeck91a, 
Moeck91b, Liu3, Mock8} and electron tomography \citep{DeRosierAndKlug, 
Cowl9, Amos19, Tamb13, Crew18, Fran14, OKeefe3D, 
Frank92, Jinschek05}, 3D information 
is being inferred from slices of 2D projections.  As followup, a recent 
direct determination of lattice parameters in three 
dimensions from images of a nanocrystal 
at two tilts \citep{wqpf.lptt} suggested that such lattice-only 
analysis strategies could be elegantly visualized
using fringe-visibility maps that were quite sensitive 
to the effects of crystal thickness. 

In this paper we investigate visibility, versus orientation, of lattice planes which 
show up as fringes due to scattering contrast effects when viewed 
nearly edge-on.   Such fringe contrast occurs for example in electron 
microscopes under both coherent (phase-contrast) and incoherent (z-contrast) 
imaging conditions.  In this analysis, we concentrate specifically on phase-contrast 
mechanisms.  

Concepts of visibility band, and band maps, are first introduced. 
Bandwidths are proportional to d-spacing over thickness, rather than 
to wavelength over d-spacing as in the case of Kikuchi \citep{nishikawa28}, 
bend-contour \citep{Hirsch77, Reimer97}, electron-channeling pattern \citep{Joy74}, 
and backscatter-electron diffraction \citep{Venables73} bands in thicker specimens.
Application examples include: 1) image-based protocols for acquisition of 
3 linearly-independent lattice plane normals from 
a single randomly-oriented nanoparticle, each parallel to an operating 
reflection g-vector, 2) predicting fringe and 
cross-fringe abundances in a collection of randomly-oriented 
particles for comparison to experimental data, and 3) measuring 
local specimen thickness by observing variations in fringe visibility during tilt.

\section{Methods}
\label{Methods}

The transmission electron microscope (TEM) used is a Philips EM430ST with a point resolution just under 0.2 nm.  Although lattice-fringe information is often 
available in electron phase-constrast images out to the ``damping limit'', 
for simplicity we only consider fringes out to the continuous-transfer or 
``spherical aberration'' limit {\em for a given image} 
since the absence of smaller fringes may be due to zeros in contrast-transfer.  
Hence talk here of ``point-resolution'' limits is shorthand for saying 
that deductions here consider only fringes whose presence in the 
sub-specimen electron-wavefield is reliably transferred to the recorded image.
Data from two samples were used for experimental measurements of the probability 
of visualizing $\left\langle 001\right\rangle$ zone fringes.  The first contained Au/Pd nanocrystals, which were sputtered onto a thin carbon film with a Hummer IV Sputter Coater. The second sample was a nanocrystalline tungsten carbide thin film, deposited by plasma-enhanced chemical vapor deposition (PECVD).

\section{Observations}

\subsection{Fringe-visibility bands}

%\begin{turnpage}
%\begin{figure*}
\begin{figure}
\includegraphics[scale=0.617]{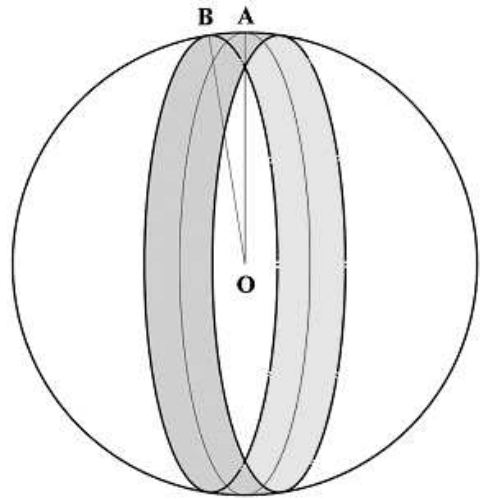}%
%\vspace{30mm}
\caption{ 
The visibility band (shaded) of a set of lattice planes of a spherical crystal.  Such a band is so defined on a sphere that when the electron beam direction lies in it, the lattice fringes are visible.  The great circle running through A is the trace of the lattice plane set.  The trace runs through the middle of a visibility band, i.e. the visibility band is symmetric about the trace.  The lattice fringes disappear as the electron beam is tilted from any point along the trace in a direction perpendicular to the trace circle by an angle greater than angle $AOB = \alpha_{max}$, the visibility band halfwidth.}
\label{Fig1}
\end{figure}
%\end{figure*}
%\end{turnpage}

On tilting away from the edge-on view of a lattice-plane, one encounters a range of incident electron angles (e.g. relative to lattice-plane parallel) within which the lattice plane's reciprocal lattice spots continue to intersect the Ewald sphere \citep{Ewald17}.  Hence lattice fringes associated with those planes remain visible. As shown in Appendix \ref{AppxB}, the upper bound of this range (with the largest term first in 
our ``thin specimen'' limit) is
\begin{equation}
\alpha_{max} = \sin^{-1} \left[ \frac{d f}{t} + \frac{\lambda}{2 d} \left( 1 - 
\left( \frac{d f}{t} \right)^2 \right) \right] .
\label{AlphaMax}
\end{equation}
Here $d$ is the spacing of lattice planes, $t$ is the crystal thickness, $\lambda$ is the wavelength of the electrons, and $f$ is a ``visibility factor'' on the order of 1 that empirically accounts for the signal-to-noise in the method used to detect fringes.  The effective radius of the reciprocal lattice spot in this model is $f/t$.  For 
example, when $f=1$ this is the first zero in $\sin[\pi g t]/(\pi g t)$: 
the signal-processing Fourier-transform of a unit-area boxcar function 
%\citep{BoxCar04}
that, convolved with the transform of an infinite lattice, 
yields the transform 
of a finite lattice of thickness $t$.  The calculation of $\alpha_{max}$ is based on kinematic conditions. This is also reasonable, since conditions become kinematic when the diffracted beam is about to extinguish.

For much of this paper, we assume that lattice fringes are visible when the electron beam lies parallel to a set of lattice planes.  As shown in Appendix \ref{AppxC}, for thick crystals under parallel illumination this may not be the case for $\alpha$ less than some $\alpha_{min}$.  Our assumption that $\alpha_{min} = 0$ is typically valid for crystals in the 10[nm] and smaller size range.  

Two symmetries shall be taken into account when considering lattice fringe visibility versus the incidence direction. The symmetries are: an azimuthal symmetry of the electron beam about the normal of the lattice planes, and a mirror symmetry of the electron beam about the lattice planes.  This consideration leads to the concept of fringe visibility band.

Such a visibility band is a schematic representation of an ensemble of electron beam incident directions.  These directions are so defined that when the electron beam is along any of them, the lattice fringes are visible. As shown in Figure \ref{Fig1},
every point on the surface of the sphere simply and elegantly represents a radially inward direction. The ensemble of orientations from which a set of lattice planes is visible form a visibility band on the surface of the sphere. The projection of a lattice plane itself is a great circle, about which the visibility band is symmetric.

%\begin{turnpage}
%\begin{figure*}
\begin{figure}
\includegraphics[scale=0.576]{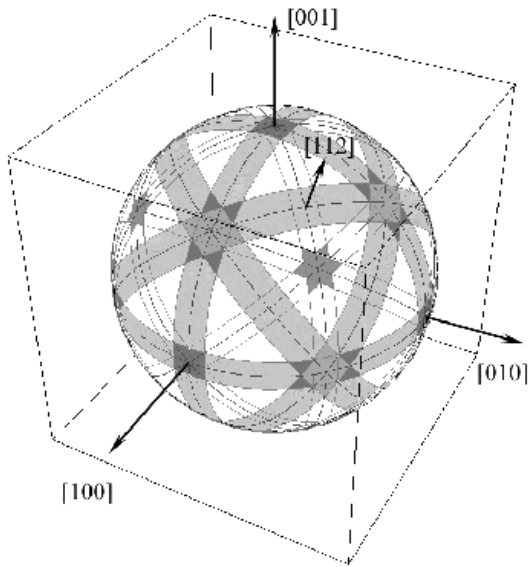}%
%\vspace{30mm}
\caption{A fringe-visibility map for a spherical fcc crystal.  Each band width is approximately proportional to the corresponding lattice spacing and to the reciprocal lattice spot size.  Like Kikuchi maps, fringe-visibility maps reflect crystal symmetry.  The latter are of special interest here as a "roadmaps" to guide crystallographic analyses in direct space.  For a randomly oriented crystal, the solid-angle subtended by each band is proportional to the probability for the corresponding lattice fringes to be visible, and the solid angle subtended by the cross-section of any two bands is proportional to the probability to get a corresponding zone-axis image.}
\label{Fig2}
\end{figure}
%\end{figure*}
%\end{turnpage

For thin specimens and the small $\lambda/d$ of typical electron microscopes, equation \ref{AlphaMax} simplifies to $\alpha_{max} \cong  \sin^{-1}(df/t) \cong df/t \propto d$.  Therefore visibility bands are different from Kikuchi bands in that the band-width is proportional (rather than inversely-proportional) to lattice spacing.  Thus we think of visibility bands as tools of ``direct space crystallography''.

\subsection{Fringe-visibility maps}

The ensemble of all the visibility bands of a spherical crystal forms a 
fringe-visibility map.  The band map not only reveals the crystal lattice 
symmetry and spacing, but also is TEM-specific, i.e. only resolvable lattice 
plane sets have their bands on the map.  Figure \ref{Fig2} shows such a map.  
Some examples to appreciate the crystallographic information in 
fringe-visibility maps are given as follows.

In the figure, four crystal directions are marked.  For this cubic lattice, the band perpendicular to the crystal direction of $[010]$ is that of the $(020)$ lattice planes, and that perpendicular to $[001]$ is the band of the $(002)$ lattice planes, etc.  The map contains bands of $\{111\}$, $\{002\}$ and $\{220\}$ lattice planes, with those of the first two classes of lattice plane sets drawn as shaded.  The absence of other bands in Fig. \ref{Fig2} suggests that the smallest lattice spacing reliably imaged by the TEM in this model is $d_{220}$.

\subsection{Applications}

\begin{figure*}
%\begin{figure}
\includegraphics[scale=0.617]{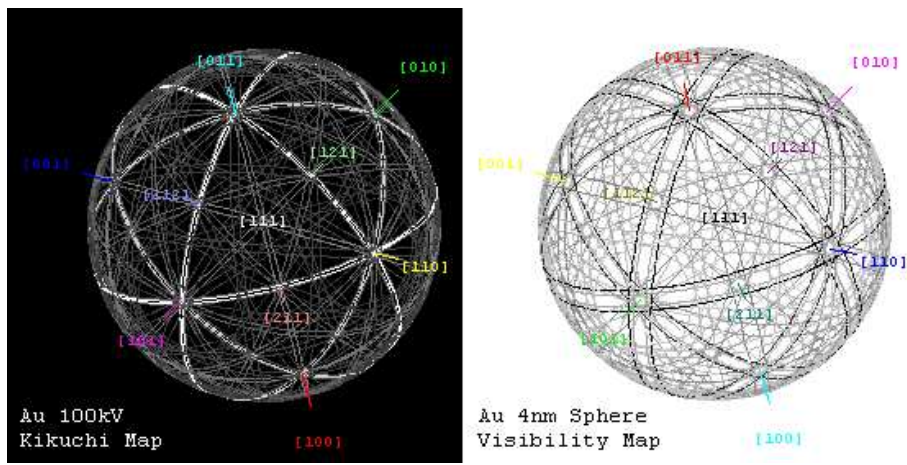}%
%\vspace{30mm}
\caption{Comparison of Kikuchi map (left) to a fringe-visibility maps (right) 
for a $t=4$[nm] single-crystal gold sphere imaged at 100[kV].  To first order, band-widths in the former (where $t \gg t_{crit}$) are proportional to $\lambda / d$, while band-widths in the latter (where $t \ll t_{crit}$) are proportional to $d / t$ .  Band boldness and d-spacing are correlated.}
\label{Fig2b}
%\end{figure}
\end{figure*}

 As shown in Fig. \ref{Fig2b}, Kikuchi maps depend primarily on electron wavelength while fringe-visibility maps are affected primarily by specimen thickness.  Thanks to the reciprocal nature of diffraction, applications of Kikuchi maps to reciprocal-space exploration are well-known.  Crystallographic information in direct-space images is increasingly available.  Hence uses for fringe-visibility maps are only beginning to emerge.  We discuss only a few here, namely determination of lattice parameters, local measurement of specimen thickness, and ``fringe'' or ``cross-fringe'' fingerprinting of randomly-oriented particle collections.  A collection of 
web-interactive maps, using the Mathematica-based Live3D applet by Martin Kraus 
\citep{Kraus03}, is available here \citep{vmlink}.

%\begin{turnpage}
%\end{turnpage

\subsubsection{Single-particle lattice parameters}

An efficient way to determine lattice parameters is by acquisition of three lattice vectors whose linear combinations span the entire reciprocal lattice \citep{Phil7}.  
Two of the three vectors may be inferred from a single zone-axis image.  For 
example \citep{wqpf.lptt} an f.c.c. WC$_{1-x}$ (a = 4.248Å) nanocrystal was 
examined with a Gatan double-tilt holder ($\pm 15^\circ$ around side-entry goniometer tilt axis, 
and $\pm 10^\circ$ around the second tilt axis).  The fringe-visibility map 
(reliably seen fringes) involved only the shaded \{111\} 
and \{002\} bands in Fig. \ref{Fig2}.  Lattice parameters 
were determined directly from $(200)$, $(020)$ and 
$(11\bar 1)$ fringes seen down [001] and [112] zone-axis directions.  

This acquisition protocol required a total tilt range of $35.3^\circ$.  The 
maximum tilt achievable with the double-tilt holder is $35.6^\circ$, barely 
higher than the tilt required.  Because of this limitation 
only particles at one goniometer extreme, with the correct 
azimuthal orientation, were candidates for the experiment.  
Specifically, the [001] zone-axis image was identified from a crystal, among thousands, at the holder setting of $(15^\circ, 9.7^\circ)$. This crystal's [001] zone 
was azimuthally oriented so that the [112] zone axis was obtained after 
tilting to $(-15^\circ, -9.7^\circ)$.

\begin{figure}
\includegraphics[scale=1.152]{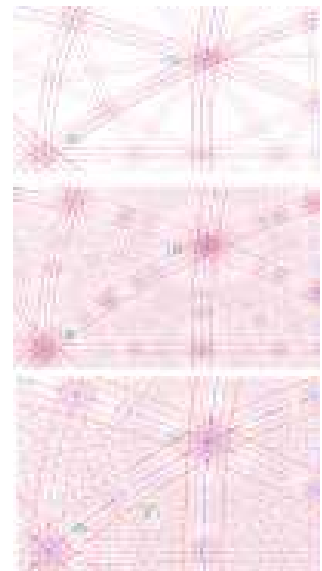} %small images
%\vspace{30mm}
\caption{Visibility band map sections for an $80\rm{\AA}$ Al-crystal at a point resolution of (a) $1\rm{\AA}$, (b) $0.6\rm{\AA}$, and (c) $30\rm{\AA}$ Al-crystal at a point-resolution of $0.6\rm{\AA}$. Improvement in point-resolution results in an increase of the number of resolvable lattice planes, and hence that of visibility bands (from a to b).  Reduction of crystal size enables visualizing lattice planes over a wider range of angles, so that the width of the bands gets larger (from b to c).
}
\label{Fig3}
\end{figure}

Thus only a small subset of particles using one specific protocol were 
eligible for this measurement.  Improvements in microscope resolution and 
goniometer range are in the works.  Recent progress in resolution enhancement 
by aberration-correction opens up myriad opportunities for image-based 
nanocrystallography \citep{Haider98, Nellist98, Aert04}.  The 
number of protocols increases at least quadratically with the microscope 
point-resolution \citep{Moeck04}. Accordingly, the fringe-visibility band 
map will contain more visibility bands and band intersections.  
Fig. \ref{Fig3} (a) and (b) show fractions of visibility band maps for an 
$80\rm{\AA}$ Al-crystal at a point resolution of $1\rm{\AA}$, 
and a $30\rm{\AA}$ Al-crystal 
at a point-resolution of $0.6\rm{\AA}$, respectively. It is clear that an 
improvement of continuous contrast transfer results in an increase of the 
number and width of visibility bands encountered. 
Thus the choice of protocols to use for lattice parameter determination 
will multiply, even were the available tilt range to remain fixed.  
The modified optics of aberration-corrected microscopes will also 
allow room for a wider range of specimen tilts.  
%\begin{turnpage}
%\begin{figure*}
%\end{figure*}
%\end{turnpage

The increase in possible analysis strategies might therefore 
overwhelm one's hope to ``make sense of all those fringes'' 
found on tilting of a randomly-oriented unknown crystal encountered 
during microscope investigation.  As shown in Fig. \ref{Fig3} (c), 
this is especially problematic if the specimen is very thin.  We therefore 
propose a protocol based on properties of ``a fringe-visibility map 
revealed piecemeal'', as one begins to examine fringes seen in 
the candidate unknown under initially small exploratory tilts \citep{Qin03}.

%\begin{turnpage}
%\begin{figure*}
\begin{figure}
\includegraphics[scale=1.188]{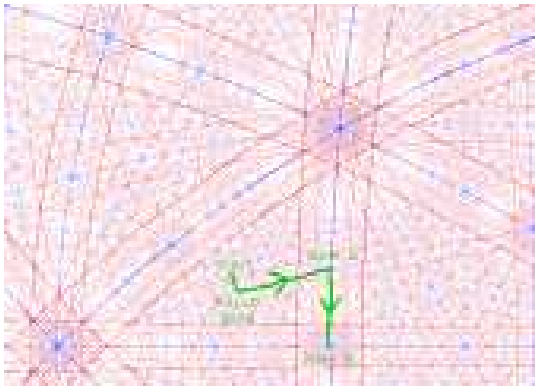} %small images
%\vspace{30mm}
\caption{Visibility band map section at $0.6\rm{\AA}$ resolution, showing a 
trajectory for acquiring three basis fringe (reciprocal lattice) vectors 
from a randomly oriented unknown.  Minimizing beam-direction errors 
in the measured lattice parameters mandates that as wide a range of 
tilts as possible be used in the determination.
}
\label{Fig4}
\end{figure}
%\end{figure*}
%\end{turnpage

The approach is based on one developed for identification 
of $100$[nm] mineral crystallites in an unequilibrated mix 
by selected area diffraction \citep{Fraundorf81}. 
It may be broken down into the following steps:

\begin{enumerate}
\item Begin with a crystal orientation at which the nanocrystal shows a set of fringes.  
If no fringes are visible, then the crystal is to be tilted randomly until fringes appear.  For example, imagine beginning at the point marked ``start'' in the fringe visibility band map shown in Figure \ref{Fig4}. 
\item	Tilt the crystal around an axis parallel to the fringes, until the center 
of the visibility band is located.
\item	Tilt the crystal around an axis perpendicular to the fringes, until a cross-fringe intersection (labeled ``Zone A'' in the map) with another band is encountered.  
\item	Similarly tilt the crystal around an axis parallel to the second band, until 
the center of zone A is located.  Record lattice spacings and goniometer settings.
\item	Tilt the crystal around an axis perpendicular to the second band, until a second cross-fringe intersection (labeled ``Zone B'' in the map) with the third band is encountered.
\item	Repeat the work described in (4) for the third band, to position the beam at the center of that zone and record it's positions and lattice spacings as well.  Calculate an oriented-basis-triplet i.e. scope coordinates of a randomly-chosen lattice basis \citep{Phil7}. 
\item	Repeat (5)-(6) to include zones at increasingly larger tilt angles 
as needed to improve accuracy of the oriented-basis-triplet determination.
\end{enumerate}

The basis triplet may at this point be reduced to conventional 
form, assuming that measurement errors are small enough.  
The precision of lattice parameter determination can be 
improved incrementally by measurement of more lattice fringes 
over a wider range of tilts.  Best-fit parameters are easily 
updated in least-squares fashion as a new set of fringes appear, 
and each fringe past the first three allows one to estimate
the precision in each direction quantitatively.  For example, 
diffraction studies indicate that lattice parameter uncertainties 
in the beam direction are initially largest, and most dependent 
on incorporation of spacings observed over a range of tilts \citep{Phil7}.  

This protocol depends on one's ability to 
tilt precisely by small angles in user-chosen directions.  
Modern double-axis goniometers can do this 
using computer support, with two caveats:  (i) mechanical 
hysteresis leaves one with a mismatch 
between goniometer reading and the actual tilt, and (ii) 
the specimen will move laterally on the nanometer scale.  
The microscopist can 
generally solve the second problem by translating to keep 
the nano-crystal of interest in the field of view.  The 
problem of tilt hysteresis is more fundamental, and will 
likely have to be addressed by independent feedback to 
verify stage orientation at a given instant.  Quantitative metrology 
with scanning probe microscopes requires independent 
verification of the probe's lateral position, instead of orientation, 
but the technologies used there (e.g. optical or capacitive 
feedback sensing) may be relevant in both cases.

\subsubsection{Random particle fringe-fingerprints}

For a randomly-oriented nanocrystal, the probability that a set of lattice 
fringes will be visible is simply proportional to the solid angle 
subtended by the corresponding fringe visibility band.  The probability 
of seeing a pair of cross-fringes is proportional to the solid angle 
subtended by the corresponding visibility-band intersections.  For 
example, for the randomly oriented spherical f.c.c. nanocrystal whose 
visibility band map is shown in Figure \ref{Fig2}, the probability of seeing 
(020) lattice planes is equal to the fraction of the $4 \pi$ solid angle 
subtended by the (020) visibility band.  For cross-fringes along the 
[001] zone-axis, the probability is equal to the fraction of all solid 
angles subtended by the cross-section of the two visibility bands 
centered on the [001] direction. 

\begin{table}
\caption{Fringe/Zone probabilities for fcc $a = 4\rm{\AA}$; $t = 50\rm{\AA}$}
\begin{tabular}{c|cc|c|c}
    & $g_{111}$ & $g_{200}$ & $\frac{p_{uvw}}{zone}$ & \# zones   \\
\hline
$\langle 110\rangle$ & 2         &  1        &     .003      &  6    \\
$\langle 100\rangle$ & 0         &  2        &     .001      &  3    \\
\hline
$\frac{p_{hkl}}{band}$  &  .046    &  .040   &    &   \\
\hline
\# bands &   4   &  3 &    &     \\
\end{tabular}
\label{TableX}
\end{table}

Thus the probability for seeing a given fringe is simply the whole-band solid 
angle divided by $4 \pi$, or $p_{hkl} = \sin[\alpha_{hkl}]$ where $\alpha_{hkl}$ 
is $\alpha_{max}$ for \{hkl\} fringes.  As shown in 
Appendix \ref{AppxA}, the solid angle subtended by a square intersection of 
bands (from Eqn \ref{Simple}) is approximately $(2 \alpha_{max})^2$.  
Hence the probability of seeing the associated cross-fringe pair is 
$p_x \cong  2 \alpha_{max}^2 / \pi$.  More generally, for an intersection 
between a band of halfwidth $\alpha_1$ and another of halfwidth $\alpha_2$ 
at an angle of $\phi$, the cross-fringe probability is 
\begin{equation}
p_x \cong \frac{2 \alpha_1 \alpha_2}{\pi \sin[\phi]} .  
\end{equation}

Using a visibility factor of $f = 0.8$ estimated from Au/Pd nanoparticles 
evaporated on a thin carbon film, $p_x$ for the $[100]$ zone in a WC$_{1-x}$ 
specimen was predicted to be about 1/700, consistent with experimental 
observation \citep{Qin03}.  Estimation of such fringe probabilities 
in turn can be used to quantitatively fingerprint collections of 
randomly-oriented nanoparticles, according to the fringes and interspot 
angles in lattice images \citep{Fraundorf04}.

\begin{figure}
\includegraphics[scale=.694]{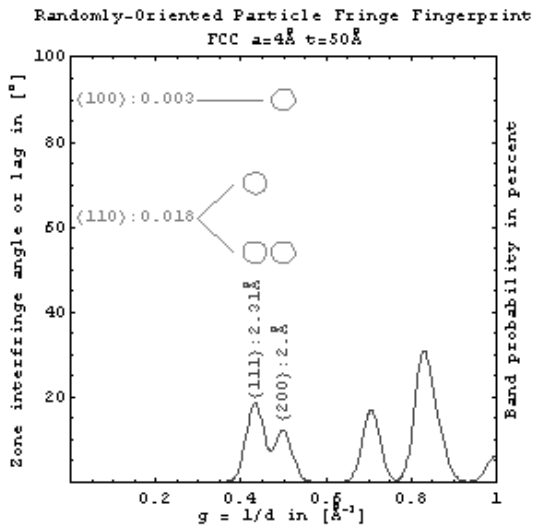}%
%\vspace{30mm}
\caption{Template for fringe-visibility fingerprinting 
of randomly-oriented fcc nano-particle collections.  Histogram
peak heights equal $p_{hkl}$, widths equal $f/t$.
}
\label{Fig4c}
\end{figure}

One can begin this process with a probability table for the 
visible fringes expected from a given type of crystal.  
For example, Table \ref{TableX} lists probabilities expected for 
the two largest lattice periodicities in a collection of $50\rm{\AA}$ fcc 
particles with a lattice parameter or $4\rm{\AA}$.  Here $p_{uvw}$ is the probability 
of encountering a fringe-pair for the [uvw] zone.  From a table 
like this, for the system of interest, one can construct a template 
like that in Fig. \ref{Fig4c} for plotting cross-fringe and 
fringe-histogram \citep{Fraundorf04b} 
data.  

Figure \ref{Fig4c} is designed to plot cross-fringe 
data points (two per spacing pair) and spacing histograms 
measured manually from multi-particle lattice fringe images.  
However, it is also patterned after the layout of a fringe 
angular covariance map \citep{Fraundorf04}.  Fluctuation 
microscopy \citep{Gibson1}, used e.g. to characterize medium-range order in 
paracrystalline specimens \citep{Gibson2}, historically has focused  
on the spacing-only or ``powder'' component 
of the pair-pair information available in a
combined spatial and angular correlation analysis of image 
data from complex materials.  The layout of Fig. \ref{Fig4c}, which 
looks at ``patches of correlated periodicity as a function of 
angular lag'' in given size range, also preserves information on 
the angle between periodicities for such fluctuation analyses.  Such plots
may therefore prove useful in the automated analysis of HRTEM images 
as well.

\begin{figure}
\includegraphics[scale=.694]{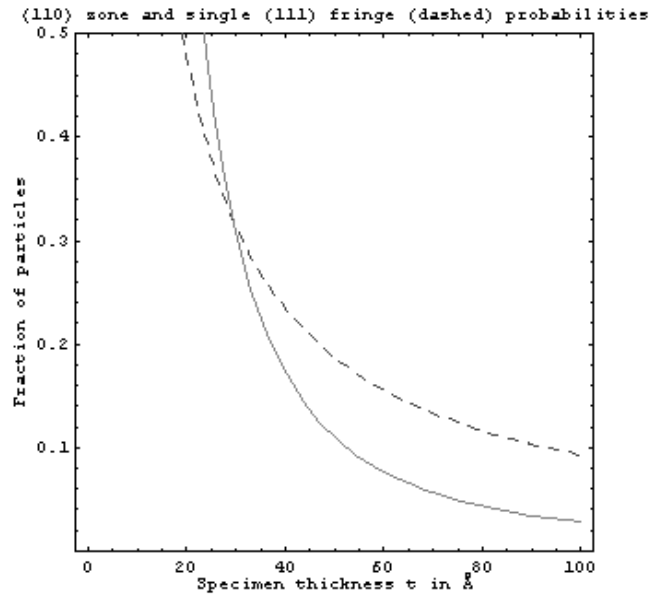}%
%\vspace{30mm}
\caption{The cross-over in abundance of 
cross-fringe versus single-fringe particles, with 
decreasing size.}
\label{Fig4d}
\end{figure}

Probability expressions flowing from fringe-visibility 
maps also allow one to quantify the relative abundance 
of particles with only one visible fringe, versus those 
with cross fringes.  For example, it nicely explains the 
fact that 2[nm] particles with cross-fringes 
are often more abundant than those with single fringes.  
This is illustrated in Fig. \ref{Fig4d}, which plots the fraction of 
particles with $\langle 110\rangle$-zone cross-fringes, versus the 
fraction with only a single \{111\}-fringe showing. 
Our observations suggest that the size at which this 
cross-over occurs is quite sensitive to viewing 
conditions, and thus to visibility factor $f$.

The precision of fringe measurements in such images \citep{Biskupek03}
is limited by shape anisotropy \citep{Warren41} and other tilt 
distortions \citep{Marks83, Spen36, Malm39}.  Nonetheless the foregoing analysis 
works in practice, at least to first order, and provides a framework 
for more careful study of these deviations as well. 

\subsubsection{Small-tilt thickness estimates}

Reduction of crystal size is well-known to result in reciprocal-space broadening of diffraction intensities as seen in X-ray, electron and neutron diffraction.  A formula, derived by Paul Scherrer for a thesis on colloids around 1915, correlates such a broadening with the inverse of the mass-weighted average grain size 
\citep{debyescherrer16}.  As shown above, decreasing size causes visibility band 
broadening as well.  Here, we discuss how band broadening can help
investigate local specimen thickness.

%\begin{turnpage}
%\begin{figure*}
\begin{figure}
\includegraphics[scale=0.899]{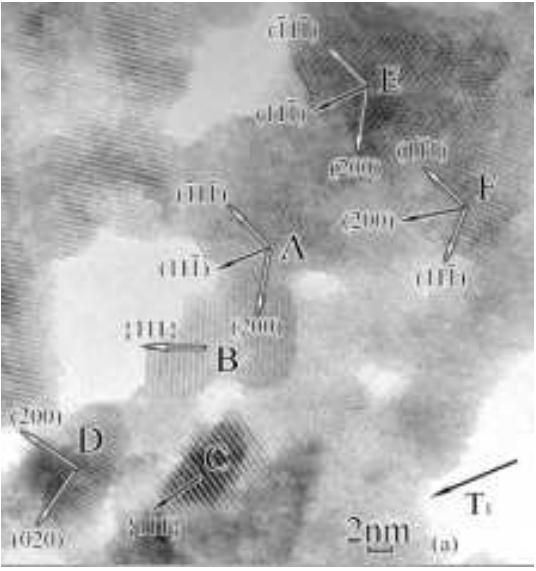} %small images
\caption{HRTEM images of six WC$_{1-x}$ nanocrystals showing lattice fringes that become invisible or remain visible after a single tilt of $14.5^{\circ}$ around the side-entry goniometer tilt axis.  The projection direction of the side-entry goniometer 
tilt axis T$_1$ is labeled at the bottom right corner.  Each 
set of parallel lattice-fringes in a nanocrystal has been labeled with both Miller indices and an arrow representing the lattice fringe normal.  The length of the arrow is proportional to 1/d.  Here, hollow arrows are used for lattice fringes that are predicted to become necessarily invisible, and solid arrows for those with certain propabilities to do so, in the second specimen orientation shown in Fig \ref{Fig5b}.}
\label{Fig5a}
\end{figure}
%\end{figure*}
%\end{turnpage}

Given a crystal's size, lattice, and its orientation with 
respect to the tilt axis, the tilt-sensitivity of lattice fringe visibilty 
can be predicted.  This is done by determining when the reciprocal lattice 
spot necessarily loses intersection with the Ewald sphere after tilt, 
given the amount of tilt used and the angle between the fringes and the tilt direction \cite{QinThesis, FringeVisibility}. In order for the reciprocal lattice spot to retain intersection with the Ewald sphere after tilt, the angular deviation of the reciprocal lattice vector from the tilt axis (or equivalently between fringe lines and the tilt direction) must be less than an upper limit $\phi$ which following Appendix \ref{AppxD} obeys:
\begin{equation}
\sin \left[\frac{\theta_{range}}{2}\right] = \frac{\sin \alpha_{max} }{\sin \phi} . 
\label{phimax}
\end{equation}
Here $\alpha_{max}$ is given by Eqn \ref{AlphaMax}.  If $\phi$ is the angle between 
an observed fringe and the tilt direction, then $\theta_{range}$ is the 
tilt-range over which that fringe is visible.  Alternatively, if $\theta_{range}$ is
an experimentally-applied tilt, then fringes whose angle to the tilt direction 
is more than $\phi$ will become invisible because their reciprocal lattice 
spot will necessarily lose intersection with the Ewald sphere after tilt.  In this 
latter application, we will refer to $\phi$ in equation \ref{phimax} as $\phi_{max}$.

%\begin{turnpage}
%\begin{figure*}
\begin{figure}
\includegraphics[scale=0.899]{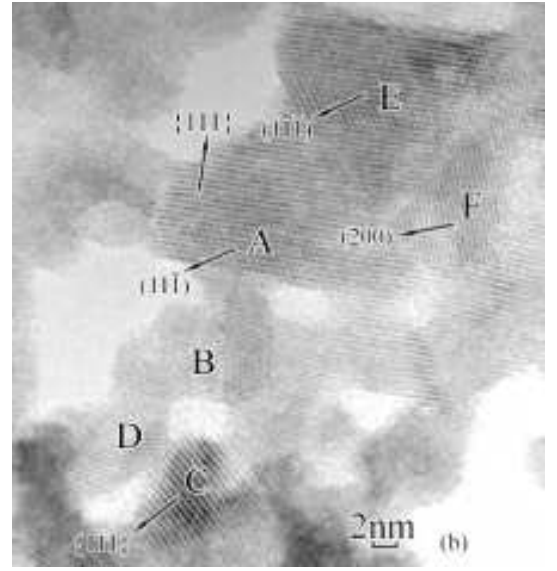} %small images
\caption{HRTEM images of the six WC$_{1-x}$ nanocrystals in Fig \ref{Fig5a}, after a single tilt of $14.5^{\circ}$ around the side-entry goniometer tilt axis.}
\label{Fig5b}
\end{figure}
%\end{figure*}
%\end{turnpage}

Equations \ref{AlphaMax} and \ref{phimax} have been used to predict lattice fringe visibility after tilt for WC$_{1-x}$ nanocrystals as shown in Figure \ref{Fig5a}. In the figure, each lattice plane set has been labeled with both the Miller indices and an arrow representing the lattice fringe normal.  The length of the arrow is proportional to 
$1/d$.  Hollow arrows are used for lattice fringes that are predicted to become necessarily invisible, and solid arrows for those with certain propabilities to do so, after tilt, as shown in the Fig \ref{Fig5b}. Take those three sets of lattice fringes of crystal A shown in Fig \ref{Fig5a}, which are those of 
the WC$_{1-x}$ (-1,1,-1), (1,1,-1) and (2,0,0) lattice planes, as examples.  The average projection size of crystal A is about $48\rm{\AA}$.  Equation \ref{phimax} predicts that for the given amount of tilt 

\begin{equation}
\phi( f = 0.79, t=48 \rm{\AA}, d_{111} = 2.453 \rm{\AA} ) = 20.6^{\circ} , \label{Limit1}
\end{equation}

and 
\begin{equation}
\phi( f = 0.79, t=48 \rm{\AA}, d_{002} = 2.124 \rm{\AA} ) = 18.3^{\circ} , \label{Limit2}
\end{equation}

The three lattice fringe normals deviate from the projection of $T_1$ by $69.1^{\circ}$, $0.3^{\circ}$, $56.6^{\circ}$, respectively.  Among them the first and the third are larger than their corresponding limits obtained in equations \ref{Limit1} and \ref{Limit2}.  Therefore, the (-1,1,-1) and (2,0,0) lattice fringes are predicted to become invisible after tilt.  This is shown to be true in Figure \ref{Fig5b}.  This way the invisibility of nine out of thirteen lattice fringe sets are predicted, which is consistent with the HRTEM observation.  The results are shown in Table \ref{Table2}.

\begin{table}
\caption{Correlating equation \ref{phimax} in predicting lattice fringe invisibility after tilt with HRTEM observation as shown in Figures \ref{Fig5a} and \ref{Fig5b}.  Quantities in the third and the fifth columns are obtained from Fig \ref{Fig5a} before tilt, and column 7 contains a theoretical prediction of invisibility after tilt for comparison to the experimental result from Fig \ref{Fig5b} in column 8.}
\begin{tabular}{cccccccc}
ID & $\theta^{\circ}$ & $t$[$\rm{\AA}$] & $(hkl)$ & $\phi^{\circ}$ & $\phi_{max}^{\circ}$ & $\frac{\phi}{\phi_{max}}>1$? & Invisible? \\
\hline
A & 14.5 & 48 & (1,1,-1)  & 0.3   & 20.6 & No  & No  \\
 &      &  & (-1,1,-1) & 69.1  &      & Yes & Yes \\
 &      &  & (2,0,0)   & 56.6  & 18.3 & Yes & Yes \\
B &      & 56 & (1,1,1)   & 24.6  & 17.8 & Yes & Yes \\
C &      & 42 & (1,1,1)   & 11.8  & 23.4 & No  & No  \\
D &      & 46 & (2,0,0)   & 57.5  & 19.0 & Yes & Yes \\
 &      &  & (0,2,0)   & 32.5  &      & Yes & Yes \\
E &      & 70 & (1,1,-1)  & 2.2   & 14.5 & No  & No  \\
 &      &  & (-1,1,-1) & 65.0  &      & Yes & Yes \\
 &      &  & (2,0,0)   & 58.2  & 13.1 & Yes & Yes \\
F &      & 48 & (1,-1,1)  & 67.1  & 20.6 & Yes & Yes \\
 &      &  & (1,1,-1)  & 43.9  &      & Yes & Yes \\
 &      &  & (2,0,0)   & 11.8  & 18.3 & No  & No  \\
\hline
\end{tabular}
\label{Table2}
\end{table}

In Figure \ref{Fig5a}, the nine lattice fringe sets which are predicted to become invisible after tilt are labeled with hollow arrows.  Please note that all these lattice fringe sets disappear in Fig \ref{Fig5b}, which is an indication of the consistency of the model with the HRTEM observations.

\begin{figure}
\includegraphics[scale=.4]{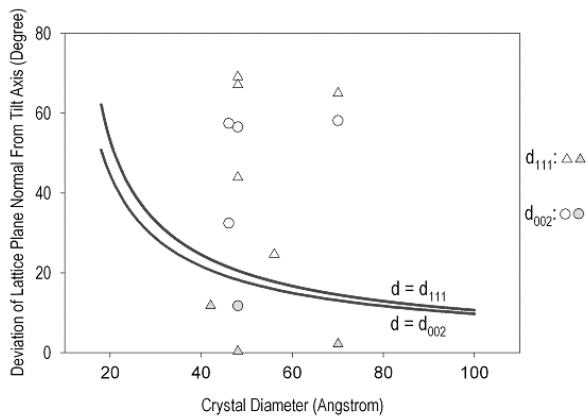}%
\caption{A plot of the maximum angular deviation of a reciprocal 
lattice vector from the tilt axis, as given by equation \ref{phimax}, 
over crystal size.   Above such a limit the reciprocal lattice spot 
necessarily loses intersection with the Ewald sphere after tilting 
the crystal by $14.5^{\circ}$.  Experimental data from Figure 
\ref{Fig5a} and \ref{Fig5b} are also shown.  The hollow symbols, 
including both circles and triangles, are used to denote the lattice 
fringe sets which are observed to become invisible after tilt, and 
solid symbols for the rest lattice fringe sets.  A consistency between 
the model and the HRTEM observation exists since all the hollow 
symbols are above their corresponding curves.}
\label{Fig6}
\end{figure}

Figure \ref{Fig6} shows plots of $\phi_{max}$ versus 
thickness with $\theta_{range} = 14.5^\circ$ and $f = 0.79$ for $d_{111}=2.453 \rm{\AA}$ 
and $d_{002}=2.124 \rm{\AA}$.  Also shown in the figure are the experimental data from Figure \ref{Fig5a}.  Hollow symbols are used to label lattice fringe sets that are observed to become invisible after tilt as shown in Figure \ref{Fig5b}, and solid symbols for the rest.  A consistency between the model and experimental observation is indicated, since all the hollow symbols are above their corresponding curves.
 
%\begin{turnpage}
%\begin{figure*}

Proof of concept in hand, one can easily solve equations \ref{AlphaMax} and \ref{phimax} for specimen thickness $t$.  Considering only the dominant term, for 
example, one gets 
\begin{equation}
t \cong  d f \csc [\frac{\theta_{range}}{2}] \csc [\phi]
\end{equation} 
where $\phi$ (as above) is the projected angle between the lattice fringe and the tilt direction.  This expression in turn suggests a fractional error in thickness, per unit error in tilt range, of 
$(1/t) \delta t / \delta{\theta} \approx t \sin[\phi] /(d f)$.  For example this predicts a 15 percent error in the thickness, per degree of error in tilt range, for a $50\rm {\AA}$ thick foil with a $2\rm {\AA}$ fringe twenty degrees from the direction of tilt.  This measurement of specimen thickness requires only verification by the microscope operator that tilt range over which the fringe remains visible is about 13 degrees.  High precision goniometers, e.g. with verifiable tilt, and a way to profile fringe visibility for small increments in tilt, could thus make it possible to routinely measure nanocrystal thickness at many points in an image with only minor amounts of tilt.  
With such instrumentation, experimental profiles of fringe intensity as 
a function of tilt like those modeled in Appendix \ref{AppxE} could open up 
a new world of real-time quantitative analysis.
%\begin{equation}
%t = \frac{{df\left( {1 + \sqrt {1 - \frac{{2\lambda \sin [\alpha ]}}{d} + 
%\left( {\frac{\lambda }{d}} \right)^2 } } \right)}}{{2\sin [\alpha ] - 
%\frac{\lambda }{d}}} \cong df\csc [\alpha ]
%\end{equation}

\section{Conclusions}

Fringe-visibility versus orientation for a set of lattice planes, relative to the incident electron beam, in specimens of given thickness is well represented by a ``visibility band''. The bands of all lattice planes reliably detected within 
the point-resolution of a microscope, during a given exposure, make up a fringe-visibility map.  Like Kikuchi-maps in reciprocal space, fringe-visibility maps serve as roadmaps for exploring orientation in direct space.  We illustrate how they can be useful for determining the 3D lattice parameters of an arbitrary nanocrystal, for fingerprinting fringe and cross-fringe abundances in a collection of randomly-oriented nanocrystals, and for determining the local thickness of crystalline specimens with modest amounts of tilt given a sufficiently precise goniometer.  Fringe-visibility maps, and these applications, will be even more useful in aberration-corrected microscopes capable of subangstrom resolution, provided attention is given (during their design) to the precise control 
{\em and verification} of specimen orientation.

\begin{acknowledgments}
This work has benefited from discussions with L. D. Marks, and 
indirectly from support by the U.S. Department of Energy and 
the Missouri Research Board, as well as by Monsanto and MEMC 
Electronic Materials Companies.
\end{acknowledgments}

\bibliography{temr1wq2.bib}

\appendix

\section{The visibility band outer half-angle}
\label{AppxB}

\begin{figure}
\includegraphics[scale=0.617]{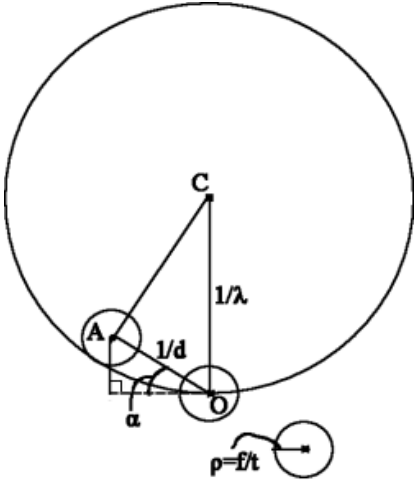}%
%\vspace{30mm}
\caption{Ewald sphere geometry for equant-particle fringe visibility.  Here
{\bf C} is the sphere's center, {\bf O} is the origin of the reciprocal 
lattice, and {\bf A} is the center of a reciprocal lattice spot.}
\label{FigB1}
\end{figure}

Figure \ref{FigB1} illustrates the geometry of a simple 
model for fringe visibility in the center of an equant (e.g. spherical) particle.  
Visibility requires that the reciprocal lattice ``spot'' 
intersect the Ewald sphere.  Point {\bf O} is the origin of the 
reciprocal lattice, and two reciprocal lattice spots are shown a 
distance $1/d$ away from the center.  Since the effective radius 
of reciprocal lattice spots is affected by many factors, it is 
written as $f/t$, where $t$ is the thickness of 
the crystal and $f$ is a visibility factor.  This visibility factor 
will be a number on the order of 1 that depends on the method of 
periodicity detection, the range of angles in the illuminating beam, 
the microscope's response function, and the amount of ``fringe 
obscuring'' noise in the field of the image.  It may need to be determined 
experimentally.  As a specimen is tilted, the reciprocal lattice spots will 
come in and out of contact with the Ewald sphere.  In Figure \ref{FigB1}, 
the outermost edge of the reciprocal lattice spot is tangent with the 
Ewald sphere defining a critical angle, $\alpha$, at which fringes of the 
corresponding spacing would be viewable in a direct-space image.  We 
derive $\alpha$ as follows.

%\begin{turnpage}
%\begin{figure*}
%\end{figure*}
%\end{turnpage}

Since the radius of the Ewald sphere is $1/\lambda$, where $\lambda$ is 
the electron wavelength, we can define the length from the center of the 
Ewald sphere {\bf C} to the center of the reciprocal lattice spot {\bf A} 
as $1/\lambda-f/t$.  The distance from the center of the Ewald 
sphere to the center of the reciprocal lattice will be $1/\lambda$ and 
the spacing between the reciprocal lattice spots will be $1/d$.  If 
$\alpha$ is the angle defining the maximum tilt of the reciprocal spot 
before loss of fringe visibility, then we can use plane trigonometry's Law of 
Cosines for the complement of angle $\alpha$ in triangle {\bf CAO}, namely 
\begin{equation}
\left( \frac{1}{\lambda} - \frac{f}{t} \right)^2 = \frac{1}{d^2}+ \frac{1}{\lambda^2} - \frac{2}{d \lambda} \cos[\frac{\pi}{2} - \alpha] ,
\end{equation} 
to obtain
\begin{equation}
sin[\alpha_{max}] = \frac{d f}{t}  + \frac{\lambda}{2 d} \left( 1 - \left( \frac{d f}{t} \right)^2 \right) . 
\label{SinAlpha}
\end{equation}
The first term (due to beam-direction broadening 
of the reciprocal lattice spot) dominates in 
HRTEM of nano-crystals because $d/t$ is typically 
greater than $\lambda/d$.  
The equation may also be relevant to 
electron-channeling, electron-backscatter diffraction, and 
(with an added factor of 1/2) Kikuchi bands.  
However, in these cases the $\lambda/d$ 
term typically dominates, yielding a 1/d rather 
than d dependence for small-angle band widths.

\section{The inner half-angle critical thickness}
\label{AppxC}

%\begin{turnpage}
%\begin{figure*}
\begin{figure}
\includegraphics[scale=0.617]{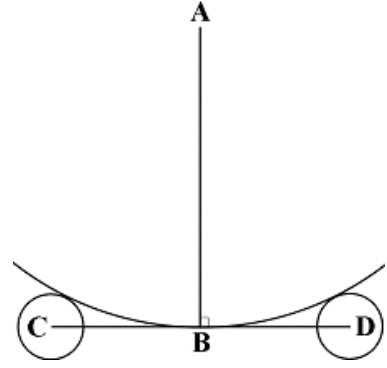}%
\caption{Schematic configuration with an electron beam traveling down a zone axis, and with reciprocal lattice spots are tangent to the Ewald sphere from the outside.  The arc centered at {\bf A} represents part of the Ewald sphere.  Segments {\bf BC} and {\bf BD} represent the reciprocal lattice vectors, i.e. $BC = g$ and $BD = -g$.  The circles centered at {\bf C} and {\bf D} represent the reciprocal lattice spots.  It is obvious {\bf CD} is perpendicular to {\bf AB}.}
\label{FigC1}
\end{figure}
%\end{figure*}
%\end{turnpage

If the specimen is sufficiently thick, or electron 
wavelengths sufficiently large, images taken with a 
parallel beam might also show loss of fringe visibility when 
the specimen is oriented on the zone axis, i.e. between 
Bragg conditions for diffraction from either side of a set of 
lattice planes.  This condition would introduce an 
``invisibility stripe" down the center of the visibility band 
depicted in Fig. \ref{Fig1}.  

Although the large Bragg angles and thick specimens 
used for X-ray diffraction make this a common occurrence, 
it is rare in electron lattice imaging because of the thin 
nature of the specimens and the small electron wavelength.  
To confirm this, consider the drawing in Fig. \ref{FigC1}.
Note that 
the length of segment {\bf AC} is $1/\lambda + f/t$, the 
length of segment {\bf BC} is $1/d$, and the length of 
segment {\bf AB} is $1/\lambda$.  Hence from 
Pythagoras' theorem for right triangle {\bf ABC}, $t$ 
becomes

\begin{equation}
t_{crit} = \frac{f d }{\sqrt{1 + (d/\lambda)^2 } - d/\lambda} .
\end{equation}

Putting in typical numbers for these quantities 
shows that this condition is seldom met for specimens 
thin enough for electron phase contrast imaging, 
particularly if effects of beam broadening (i.e. a range 
of angles in the incident beam) are 
taken into account.  For thicknesses great enough, 
simply changing the sign of $f/t$ in Eqn. \ref{SinAlpha} 
tells us that the half-width $\alpha_{min}$ 
of the invisibility stripe will obey
\begin{equation}
\sin[\alpha_{min}] = \frac{\lambda}{2 d} \left( 1 - \left( \frac{d f}{t}\right)^2 \right) - \frac{d f}{t}, 
\end{equation}
where the first term is the dominant one.

\section{Estimating band intersection areas}
\label{AppxA}

%\begin{turnpage}
%\begin{figure*}
\begin{figure}
\includegraphics[scale=0.694]{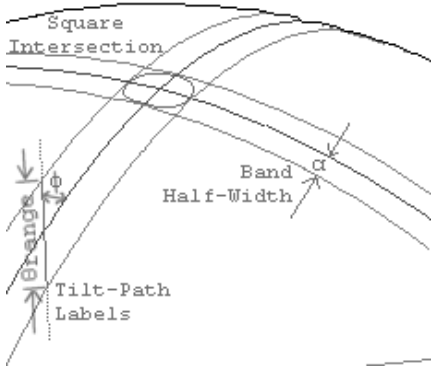}%
\caption{Visibility bands, a tilt path, and 
a ``cross-fringe'' intersection on the surface of a unit sphere.}
\label{FigA1}
\end{figure}
%\end{figure*}
%\end{turnpage

Consider first the case of equal bands intersecting at right angles, 
as shown in Fig. \ref{FigA1}.  We are interested in $\sigma_2$, twice 
the area $\sigma_1$ of one intersection, because great-circle bands 
intersect twice on opposing sides of the orientation unit sphere.  
For the inscribed circle shown in the figure, the 
``double-side solid angle'' is $4 \pi (1 - \cos \alpha)$.  
The double-side solid angle of the circumscribed circle is 
approximately $4 \pi (1- \cos (\sqrt 2 \alpha))$.

For $\alpha < \pi/4$, the {\em exact} value of the 
intersection solid angle \citep{Bailey01} is:
\begin{widetext}
\begin{equation}
\sigma_1 = 8 \sum\limits_{n = 1}^\infty  {\frac{{( - 1)^{n + 1} \sin[\alpha]^{2n} }}{{n!}}} F[n]\prod
\limits_{k = 1}^n {(\frac{3}{2} - k)} = 4 \alpha^2 + \frac{2}{9} \alpha^6 + \frac{8}{45} \alpha^8 + ...,
\end{equation}
\end{widetext}
where hypergeometric function $F[n]$ is 
\begin{equation}
F[n] \equiv \sum\limits_{m = 0}^{n-1} \frac{(n-1)!}{m!(n-1-m)!} \frac{1}{2m+1} .
\end{equation}  For $\alpha > \pi/4$, upper and lower visibility zones 
connect leaving only four circular ``cross-free caps''.  Hence $\sigma_1$ 
then becomes $2 \pi (2 \sin[\alpha] - 1)$.  

These observations show that an excellent approximation 
for small angles is $\sigma_2 \cong 2(2 \alpha)^2$, i.e. twice the area 
one would calculate for a flat square of side $2 \alpha$.  The approximation 
error is to first order $(4/9) \alpha^6$, and is still below $0.5\%$ 
when bandwidth $2 \alpha$ is a radian.

More generally, the bandwidth half-angles 
(e.g. $\alpha$) follow from the elements 
described in Appendix \ref{AppxB}, namely the 
lattice spacing, specimen thickness, and electron wavelength.  
When crossing bands have half-angles of $\alpha_1$ and 
$\alpha_2$, and intersect at 
an angle of $\phi$ radians, the flat polygon estimate becomes
\begin{equation}
\sigma_2 \cong \frac{2 (2 \alpha_1)(2 \alpha_2)}{\sin[\phi]}.
\label{Simple}
\end{equation}
Given the value (or an estimate) for visibility-band intersection 
solid-angles, the probability of seeing 
cross-fringes in a randomly-oriented particle is then (assuming 
negligible zone overlap) simply 
$p_x = n \sigma_2 / (4 \pi)$, where $n$ is the zone multiplicity 
e.g. n=3 for $\langle 100 \rangle$ zone cross-fringes from a cubic crystal.

\section{The tilt-range for fringe visibility}
\label{AppxD}

%\begin{turnpage}
%\begin{figure*}
\begin{figure}
\includegraphics[scale=0.775]{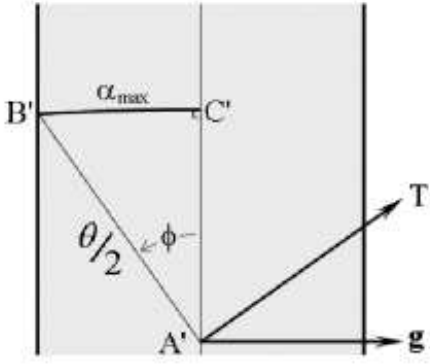} %small images
\caption{A segment of a continuous visibility band, with half 
of the visibility ``tilt range'' marked by segment A'B'.}
\label{FigD1}
\end{figure}
%\end{figure*}
%\end{turnpage

The angular range over which a set of lattice planes remains visible is smallest if the tilt axis is parallel to the planes.  Half of this angular range is then given by Eqn. \ref{AlphaMax}. Generally, the tilt axis is not parallel to the lattice planes.  To take advantage of band symmetry, we begin with the electron beam initially parallel to the lattice planes as shown in Fig. \ref{FigD1}.  The tilt range for fringe visibility is then easily quantified.

The figure displays a visibility band segment.  A'C' is the trace of the lattice plane.  A' is the starting electron beam direction.  T is the tilt axis, g is the reciprocal lattice vector.  T and g make an angle of $\phi$.  Since B'C'$\perp$A'C', B'C' = $\alpha_{max}$ is half of the visibility band width.  A'B' is the tilt path of the electron beam direction, and is half the total tilt range $\theta$ within which the lattice fringes are visible (the other half is symmetric with A'B' about A'). The following relationships are obvious from the figure: A'B'$\perp$T, $\angle$B'A'C' = $\phi$, and each of A'B', B'C', and C'A' is an arc of a great circle. Spherical trigonometry's Law of Sines then yields:
\begin{equation}
\sin [\alpha_{max}] = \sin[\phi] \sin[\theta /2] .
\end{equation}

\section{Fringe-visibility rocking curves}
\label{AppxE}

Given the shape of a nanocrystal, the 
vector separation between a reciprocal lattice point 
and the Ewald sphere allows one to calculate the Fourier 
intensity of an individual fringe as a function of 
orientation.  Replacing $f/t$ with $g_{sep}$ in 
Eqn. \ref{AlphaMax}, and solving, gives for the magnitude of 
this vector separation
\begin{equation}
g_{sep} = \left| \frac{\sqrt{1 + (\lambda /d)^2 - 2 (\lambda /d) \sin[\alpha]} - 1} {\lambda} \right|,
\end{equation}
where $\alpha$ is an arbitrary tilt of the beam direction from edge-on.

Because fringe-intensity profiles concern not just the 
boundaries of fringe visibility, dynamical 
contrast effects must be taken into account, particularly 
for crystal thicknesses near to or larger than an extinction 
distance \citep{Hirsch77, Reimer97}.  Here, however, we illustrate 
such profiles or rocking curves for the 
simplest case, namely a spherical particle of diameter $t$ 
in the kinematic approximation.  For such a particle, 
the shape transform as a function of spatial frequency 
$g$ (again using signal-processing conventions) is simply
\begin{equation}
s = \frac{\sin[\pi g t] - \pi g t \cos[\pi g t]}{2 \pi^2 g^3} .
\end{equation}

\begin{figure}
\includegraphics[scale=0.69]{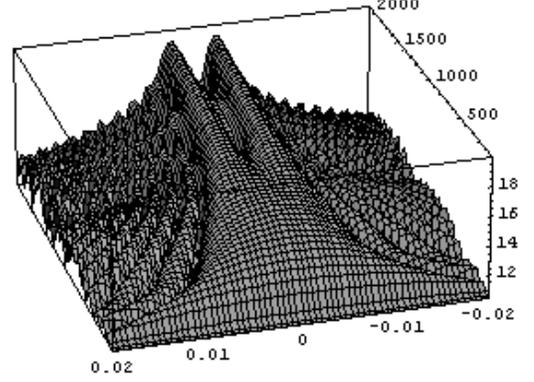}%
\caption{Logarithmic fringe-intensity profile ($d = 2\rm{\AA}$ and 
$\lambda = 0.01\rm{\AA}$) for a spherical particle whose diameter 
$t$ is given in Angstroms by the top-labeled axis (running front to back), 
as a function of ``tilt from edge-on in radians'' on the bottom-labeled 
axis (running left to right).}
\label{FigE1}
\end{figure}

Consider a lattice plane canted by $\alpha$ radians 
from the edge-on position along the electron beam 
direction.  Since all reciprocal lattice spots 
will be convolved with the shape transform, one can 
add amplitudes (in the coherent scattering case) from
both sides of the lattice plane by adding 
s-values for $g_{sep}$ evaluated at $\pm \alpha$.  
Fig. \ref{FigE1} illustrates.  

%\begin{turnpage}
%\begin{figure*}
%\end{figure*}
%\end{turnpage

Note that for small thicknesses the fringe intensity profile 
shows the expected $(d / t)$-dependence of its 
angular half-width.  For spheres of diameter greater 
than $650 \rm{\AA}$, the rocking curve bifurcates 
into the fixed $(\lambda / d)$-width band cut 
by the reduced-visibility stripe predicted in 
Appendix \ref{AppxC}.  This is therefore an alternate view of the 
transition between thick specimen electron-channeling/Kikuchi 
map ($\lambda / d$) and thin specimen fringe-visibility 
map ($d / t$) geometries.  Bend-contours of suitably-oriented 
wedge-shaped crystals thus likely (cf. Fig. 9 \citet{Hashimoto62}) 
bridge the gap experimentally between both extremes.

% Bibliographic references with the natbib package:
% Parenthetical: \citep{Bai92} produces (Bailyn 1992).
% Textual: \citet{Bai95} produces Bailyn et al. (1995).
% An affix and part of a reference:
%   \citep[e.g.][Ch. 2]{Bar76}
%   produces (e.g. Barnes et al. 1976, Ch. 2).

%\begin{thebibliography}{}

% \bibitem[Names(Year)]{label} or \bibitem[Names(Year)Long names]{label}.
% (\harvarditem{Name}{Year}{label} is also supported.)

%\bibitem[Kung and Foecke (1999)] {Kung H. and Foecke T. (1999), 
%"Mechanical behavior of nanostructured 
%materials", MRS Bulletin, 24, 14-15. "Mechanical behavior of nanostructured materials", MRS Bulletin, 24, 14-15}

%\bibitem[]{}

%\end{thebibliography}

\end{document}